\documentclass[english,aps,prl,reprint,floatfix,preprintnumbers,nofootinbib,superscriptaddress,longbibliography]{revtex4-2}
\usepackage[utf8]{inputenc}
\usepackage[english]{babel}
\usepackage{amsmath,amssymb,mathtools,bm,graphicx,booktabs,multirow}
\usepackage{slashed}
\usepackage{BOONDOX-frak}
\usepackage{microtype}
\usepackage{tikz,pgfplots}
\pgfplotsset{compat=1.18}
\usepgfplotslibrary{fillbetween}
\usetikzlibrary{arrows.meta,positioning}
\usepackage[colorlinks=true,linkcolor=blue!55!black,citecolor=blue!55!black,urlcolor=blue!55!black]{hyperref}

\allowdisplaybreaks[3]
\newcommand{\dd}{\mathrm{d}}

\newcommand{\Lag}{\mathcal{L}}
\newcommand{\hc}{\mathrm{h.c.}}
\newcommand{\nn} {\nonumber\\}

\newcommand{\Section}[1]{\noindent{\bf #1.}}

\usepackage[draft,authormarkup=none]{changes}
\definecolor{changesadded}{HTML}{6B9FD1}
\definecolor{changesdeleted}{HTML}{D98A8A}
\setaddedmarkup{\textcolor{changesadded}{#1}}
\setdeletedmarkup{\textcolor{changesdeleted}{\ifmmode\text{\sout{\ensuremath{#1}}}\else\sout{#1}\fi}}
\newcommand{\eref}[1]{Eq.~\eqref{#1}}

\begin{document}
\title{Moment bounds and UV reconstruction from curved positivity cones}
\preprint{{\footnotesize USTC-ICTS/PCFT-26-88}}

\author{Fu-Ming Chang}
\email{changfum@mail.ustc.edu.cn}
\affiliation{Interdisciplinary Center for Theoretical Study, University of Science and Technology of China, Hefei, Anhui 230026, China}
\author{Shi-Lin Wan}
\email{wsl9868@mail.ustc.edu.cn}
\affiliation{Interdisciplinary Center for Theoretical Study, University of Science and Technology of China, Hefei, Anhui 230026, China}
\author{Shuang-Yong Zhou}
\email{zhoushy@ustc.edu.cn}
\affiliation{Interdisciplinary Center for Theoretical Study, University of Science and Technology of China, Hefei, Anhui 230026, China}
\affiliation{Peng Huanwu Center for Fundamental Theory, Hefei, Anhui 230026, China}
\date{\today}

\begin{abstract}
The fundamental principles of the S-matrix impose positivity bounds on the Wilson coefficients of EFTs, including the SMEFT. These bounds form convex cones that often have curved boundaries when there are degeneracies among the irreducible two-particle low-energy states. Numerical approximations have previously been used to evaluate these curved cones. We develop a moment-based framework to analytically carve out these cones, particularly those associated with anomalous quartic gauge couplings in the SMEFT. The framework also allows us to reverse engineer UV models from the curved boundaries of the positivity cones.

\end{abstract}

\maketitle

Requiring effective field theories (EFTs) to admit consistent UV completions places non-trivial constraints on their low-energy interactions (see, eg \cite{Adams:2006sv,Tolley:2020gtv,Caron-Huot:2020cmc, Manohar:2008tc, Bellazzini:2016xrt, deRham:2017avq,deRham:2017zjm, Arkani-Hamed:2020blm, Bellazzini:2020cot, Sinha:2020win, Alberte:2020jsk, Tokuda:2020mlf, Caron-Huot:2021rmr, Guerrieri:2021tak, Alberte:2021dnj, Du:2021byy, Chiang:2021ziz, Bern:2021ppb, Eliasmiro:2022xaa, Albert:2022oes, Caron-Huot:2022ugt, Henriksson:2022oeu, Wan:2024eto, Cheung:2024uhn, Berman:2026ezk, Wan:2026pjq, Huang:2026ghg} and \cite{Derham:2022hpx} for a review). These constraints, known as positivity bounds, can be derived from dispersion relations that connect low-energy scattering amplitudes to the underlying UV dynamics, utilizing analyticity, unitarity, and crossing symmetry. 
Within the Standard Model EFT (SMEFT), a model-independent framework for describing the low-energy effects of heavy new physics \cite{Weinberg:1978kz,Leung:1984ni,Buchmuller:1985jz} (see \cite{Isidori:2023pyp} for a review), positivity bounds have been extensively studied to constrain the allowed Wilson coefficient space and extract their physical implications~\cite{Zhang:2018shp,Bellazzini:2018paj,Bi:2019phv,Remmen:2019cyz,Zhang:2020jyn,Yamashita:2020gtt,Remmen:2020vts,Fuks:2020ujk,Gu:2020ldn,Bonnefoy:2020yee,Li:2021lpe,Zhang:2021eeo,Ghosh:2022qqq,Li:2022tcz,Li:2022rag,Li:2022aby,Altmannshofer:2023bfk,Yang:2023ncf,Davighi:2023acq,Chala:2023xjy,Chen:2023bhu,Hong:2024fbl,Remmen:2024hry,Chakraborty:2024ciu,Liu:2025deo,Chang:2026ztn}.

In particular, electroweak boson scattering at colliders provides an important probe of the physics beyond the Standard Model~\cite{He:1994br,ATLAS:2020nlt,Covarelli:2021gyz}. Deviations from the SM predictions for these processes can arise from dimension-8 anomalous quartic gauge couplings (aQGCs)~\cite{Li:2020gnx, Murphy:2020rsh} (see also \cite{Remmen:2019cyz, Almeida:2020ylr, Durieux:2024zrg}), which contribute to 2-to-2 scattering amplitudes at order $s^2$ ($s$ being the Mandelstam variable).
The positivity-allowed $s^2$ coefficients form a convex cone, hereafter called the amplitude cone, which can be described by its extremal rays (ERs)~\cite{Zhang:2020jyn, Bellazzini:2014waa, Zhang:2021eeo, Remmen:2024hry, Bonnefoy:2025uzf}. Earlier studies constructed the ERs for several subsectors of the SMEFT aQGC cone~\cite{Yamashita:2020gtt, Zhang:2021eeo}. This construction has recently been extended to all dimension-8 aQGC operators, yielding the complete set of ERs~\cite{Chang:2026ztn}.

Describing the amplitude cone in terms of analytic inequalities, ie positivity bounds, would clarify the constraints on Wilson coefficients and facilitate comparisons with experimental results and candidate UV models. For a cone with finitely many ERs, these inequalities can be obtained by facet enumeration~\cite{Zhang:2020jyn, Zhang:2021eeo}. However, degeneracies among irreducible two-particle states can give rise to continuous families of ERs and curved boundaries, as occurs in sectors of the SMEFT. Obtaining exact positivity bounds from these continuous families of ERs remains challenging, as a direct approach requires optimizing high-degree polynomials over multiple parameters.

In this Letter, we develop a systematic moment-based framework to derive exact positivity bounds for curved amplitude cones and address the corresponding inverse problem of constructing possible UV models from IR data. Moment methods have been widely used to derive positivity bounds by expressing dispersive sum rules in terms of moments~\cite{Arkani-Hamed:2020blm,Bellazzini:2020cot,Chiang:2021ziz,Chiang:2022ltp,Chiang:2022jep,Wan:2024eto,Calisto:2026pvv}. Here we apply the moment approach directly to the continuous ER families. Their simple polynomial structure allows positive sums over ERs to be converted into positive semidefinite (PSD) constraints on ER-induced moments. Eliminating auxiliary moments then leads to analytical bounds directly on Wilson coefficients. We further show how the curved boundaries encode information about UV interactions. In particular, when a bound is saturated, null vectors of the moment matrices impose relations among the UV couplings to different scattering channels, which can be used to construct explicit UV Lagrangians.

\Section{Dispersion relations}
Let us start with dispersion relations, which connect low-energy EFT amplitudes to possible UV spectra about which we remain agnostic. We define $\mathcal M_{ijkl}$ as the coefficients of $s^2$ in the forward $ij\to kl$ scattering amplitudes, where $i,j,k,l$ label the external particle states. Combining analyticity, crossing symmetry, and unitarity, we obtain the following sum rules, with low-energy contributions below the EFT cutoff $\Lambda$ subtracted,
\begin{align}
 \!\!\!\!\!\mathcal M_{ijkl}
 &=\!\!\int_{\Lambda^2}^{\infty}\frac{\dd s}{2\pi s^3}
 \sum_X\! A_{ij\to X}A_{kl\to X}^{*}
 \!+\!(j\leftrightarrow l) \label{eq:multidisp}
\end{align}
where $A_{ij\to X}$ is the on-shell amplitude for the transition from the two-particle state $(ij)$ to the intermediate state $X$ and $(j\!\leftrightarrow\!l)$ denotes the $su$ crossed term. 
In the SMEFT, for simplicity, we shall focus on the $s^2$ contributions from dimension-8 aQGC operators, since dimension-6 effects shift $\mathcal M_{ijkl}$ without changing the structure of the subsequent analysis. (The dimension-6 effects always contribute negatively to the $s^2$ positivity bounds, so the dimension-8 only bounds are conservative ones \cite{Chang:xxx}.)

The sum rules \eqref{eq:multidisp} imply that $\mathcal{M}_{ijkl}$ lies within a convex cone generated by its ERs \cite{Zhang:2020jyn}.
For notational convenience, we represent $\mathcal{M}_{ijkl}$ as a vector $\mathbf{F}$ in the amplitude space. Denoting the cone's ERs as $\mathbf{e}_a$, we have 
\begin{align}
\!\!\mathbf{F}&=\sum_{a}w_a \mathbf{e}_a+\sum_{b} \!\int\! w_{b}(\mathbf{z}_b)\,\mathbf{e}_{b}(\mathbf{z}_{b})\dd \mathbf{z}_{b},~~w_a, w_{b}\geq0,
\label{eq:weightedgenerators}
\end{align}
where $a$ labels the discrete ERs, $b$ labels the continuous ER families, and $\mathbf{z}_b=(z_{b,1}, z_{b,2},...)^T$ denotes the corresponding continuous parameters arising from possible degeneracies among two-particle channels \cite{Chang:2026ztn}. These ERs can be constructed from the underlying symmetries without specifying the detailed UV dynamics.
However, the ER representation of the amplitude cone does not directly yield explicit inequalities on the Wilson coefficients. For continuous families of ERs, previous studies have used discretization to approximate the amplitude cone by a polyhedral cone and evaluate positivity bounds, particularly when the ERs depend on several continuous parameters.

\Section{Moment approach}
We now propose a moment approach that treats continuous families of ERs collectively, providing a systematic route to analytic constraints while avoiding computationally expensive discretization of their parameters.
The key observation is that, due to the monomial nature of the continuous ER families $\mathbf e_c(\mathbf z_c)=\sum_{\alpha,\beta}\boldsymbol\rho_c^{\alpha\beta}\mathbf z_c^\alpha\mathbf z_c^\beta$ with $\mathbf z_c^\alpha\equiv \prod_{i=1}^{n_c}z_{c,i}^{\alpha_i}$, \eref{eq:weightedgenerators} can be cast as a sum of multivariate moments $\mathtt{m}^{\alpha\beta}_{c}$,
\begin{align}
  \mathbf{F}&=\sum_{c}\int\mathbf{e}_{c}(\mathbf z_c)\,\dd\mu_{c}(\mathbf z_c),~~~~~~\dd\mu_{c}(\mathbf{z}_c)\geq 0\nonumber\\
  &=\sum_{c,\alpha,\beta} \boldsymbol{\rho}^{\alpha\beta}_{c}\,\mathtt{m}^{\alpha\beta}_{c},~~~~\mathtt{m}^{\alpha\beta}_{c}=\int \mathbf{z}^\alpha_{c}\mathbf{z}^\beta_{c}\dd\mu_{c}(\mathbf{z}_c), \label{eq:summomF}
\end{align}
where the sum over $c$ includes the sums over $a$ and $b$, $\dd\mu_{a}(\mathbf{z}_a)= w_a\delta(\mathbf{z}_a)\dd \mathbf{z}_a$ and $\dd\mu_{b}(\mathbf{z}_b)=w_{b}(\mathbf{z}_b)\,\dd \mathbf{z}_b$. Since the continuous parameters $\mathbf z_c$ range over real spaces, $\mathtt{m}^{\alpha\beta}_{c}$ are multivariate Hamburger moments, whose analytical solution is well known. 

In general, the existence of a moment integral representation is equivalent to some conditions on the moments. For a truncated Hamburger problem with $\alpha+\beta\le d_c$, the Hankel matrix condition $\mathcal H_c\succeq0$, with $(\mathcal H_c)_{\alpha\beta}=\mathtt m_c^{\alpha\beta}$, provides a necessary one~\cite{Wan:2024eto}. For the homogeneous quadratic ER families plus discrete ERs considered here, including those of the SMEFT aQGC cone, it is actually sufficient to impose PSD on the block-diagonal Hankel matrices $\mathcal H_c=\operatorname{diag}(\mathtt m_c^{00},\hat{\mathcal H}_c)$, since the off-diagonal components drop out of the sum rule \eqref{eq:summomF}.

Therefore, if it exists, the optimal lower bound on a linear combination of Wilson coefficients, ${\rm Obj}({\bf F})$, can be obtained from the following semidefinite program:
\begin{align}
\underset{\mathtt m}{\operatorname{minimize}:}~~
&{\rm Obj}({\bf F})\\
\text{subject to:}~~
&\mathbf F=\!\sum_{c,\alpha,\beta}\! \boldsymbol{\rho}^{\alpha\beta}_{c}\,\mathtt{m}^{\alpha\beta}_{c},~~~~\mathcal H_c(\mathtt m)\succeq0~~\forall c ,\nonumber
\end{align}
while the upper bound can be obtained by minimizing $-{\rm Obj}({\bf F})$. Bounds on a slice of the amplitude cone normal to $\mathbf r$ follow by adding the constraint $\mathbf r^T\mathbf F=1$.

Let us illustrate how the PSD Hankel matrix constrains the parameter space with a toy example with just one continuous ER family: $\bm e(x,y)=(x^2,2xy,y^2)$. In this case, $\bm F\equiv (C_1,C_2,C_3)=(\mathtt{m}^{20},2\mathtt{m}^{11},\mathtt{m}^{02})$ is subject to the following Hankel matrix condition,
\begin{equation}
 \hat{\mathcal{H}}
  =\begin{pmatrix}
    \mathtt{m}^{20} &\mathtt{m}^{11} \\
    \mathtt{m}^{11} &\mathtt{m}^{02} 
  \end{pmatrix} =\int
  \begin{pmatrix} z_1 \\ z_2 \end{pmatrix}
  \begin{pmatrix} z_1 & z_2 \end{pmatrix}\dd\mu
  \succeq0 .
\label{eq:psdtoyH}
\end{equation}
This matrix is PSD because $\bm v^T\hat{\mathcal H}\bm v=\int(v_1 z_1+v_2 z_2)^2\,\dd\mu\geq0$ for any vector $\bm v$, as $\mu$ is a non-negative measure.
It implies $C_1,C_3\geq0$ and $C_2^2\leq 4C_1C_3$, with the latter describing a curved boundary that reflects a correlation between different couplings.

\Section{Quantifier elimination}
In general, however, one cannot use \eref{eq:summomF} to express all the relevant moments in $\mathcal H_c$ in terms of the amplitude coefficients in $\mathbf F$. Therefore, the conditions $\mathcal H_c\succeq0$, which are equivalent to requiring all their principal minors to be non-negative, lead to polynomial inequalities in the Wilson coefficients and some auxiliary moments. Note that these $\mathcal H_c$ matrices are usually small. In the aQGC case, $\hat{\mathcal H}_c$ in $\mathcal H_c=\operatorname{diag}(\mathtt m_c^{00},\hat{\mathcal H}_c)$ is at most $4\times 4$. Obtaining bounds solely on the Wilson coefficients requires projecting the feasible region onto coefficient space by eliminating these auxiliary variables, which is a standard problem of quantifier elimination \cite{Basu:2006}.

In its simplest form, we can carry out this elimination by determining the allowed range of an auxiliary moment under each constraint and requiring these ranges to have a nonempty intersection. For a higher-degree polynomial inequality, this intersection condition is equivalent to requiring the polynomial's maximum over the interval allowed by the remaining constraints to be non-negative. We can also eliminate free off-diagonal auxiliary moments using chordal completion, replacing large PSD constraints with equivalent conditions on smaller matrix blocks~\cite{Grone:1984}. In what follows, we shall illustrate how this can be done with a few examples.

\Section{Bounds in bi-scalar EFT} 
First, consider a generic bi-scalar theory, where the amplitude vector is $\bm F\equiv(C_1,\ldots,C_6)$ and the convex cone is generated by the ERs \cite{Zhang:2021eeo}
\begin{align}
(z_1^2,2z_1z_2,z_2^2,4z_3^2,4z_1z_3,4z_2z_3)~\text{and}~(0,-1,0,1,0,0)\nonumber.
\end{align}
The allowed parameter space is determined by the existence of $\mathtt{a}\geq0$ such that
\begin{align}
 \!\!\!\!\hat{\mathcal H}(\mathtt{a})=\begin{pmatrix}
 C_1&(C_2+\mathtt{a})/2&C_5/4\\
 (C_2+\mathtt{a})/2&C_3&C_6/4\\
 C_5/4&C_6/4&(C_4-\mathtt{a})/4
 \end{pmatrix}\succeq0,
 \label{eq:threechannel}
\end{align}
where $\mathtt{a}$ is the $00$ moment of the discrete ER. 
Focusing on the diagonal entries, we have $C_1,C_3,C_4\geq0$ because $\mathtt{a}\geq0$. Denoting $\tilde{C}_1=2\sqrt{C_1C_3}$, all $1\times1$ and $2\times2$ principal minors being non-negative restrict $\mathtt{a}$ to $[\ell,u]$, where
\begin{align}
 \ell(\bm F)&=\max(0,-C_2-\tilde{C}_1),\\
 u(\bm F)&=\min\Big(C_4,-C_2+\tilde{C}_1,C_4-\frac{C_5^2}{4C_1},
 C_4-\frac{C_6^2}{4C_3}\Big)\nonumber.
\end{align}
The remaining condition is that the $3\times3$ principal minor is non-negative, $\det \hat{\mathcal{H}}(\mathtt{a})\geq 0$. 
Its overlap with $\ell \leq \mathtt{a}\leq u$
yields the analytic bounds:
\begin{align}
 \ell(\bm F)&\leq u(\bm F),\nonumber\\
 (C_4+C_2-\tilde{C}_2)\left(\tilde{C}_1^2-\tilde{C}_2^2\right)&
 +\tilde{C}_2C_5C_6\geq C_1C_6^2+C_3C_5^2.\nonumber
\end{align}
where $\tilde{C}_2=C_2+\min \{u,\max[\ell, (C_4-2C_2-\sqrt{\Delta})/3]\}$ and $\Delta=\max [0,(C_2+C_4)^2+12C_1C_3-3C_5C_6]$, in agreement with Ref~\cite{Li:2021lpe}.

\Section{Bounds in $B$-$H$ sector} \label{sec:BH}
Next, we turn to the SMEFT sector involving the hypercharge gauge boson $B$ and the Higgs doublet $H$. In the amplitude basis of the relevant dimension-8 aQGC operators, the $s^2$ amplitude vector is 
\begin{align}
 \bm F=(F_{M2},F_{M3},F_{S0},F_{S1},F_{S2},F_{T8},F_{T9}),
\end{align}
where $F_{i}$ are the standard dimensionless Wilson coefficients; see Ref~\cite{Durieux:2024zrg, Chang:2026ztn} for more details. 
The UV symmetries yield eight discrete ERs,
\begin{alignat}{3}
  &(0,0,2,0,-2,0,0),(0,0,0,0,0,-2,4),(0,0,0,0,0,-1,4)\nonumber\\
  &(0,0,0,-2,2,0,0),(0,0,2,-4,2,0,0),(0,-1,0,0,0,0,0)\nonumber\\
  &(0,0,2,0,0,0,0),(-1,-2,0,0,0,0,0)
  \label{eq:BHERs}
\end{alignat}
and one continuous family of ERs, 
\begin{align}
\bm e_{\rm BH}(z_1,z_2)=(4z_1z_2,0,0,2z_2^2,0,2z_1^2,0).
 \label{eq:BHfamily}
\end{align}
We assign a non-negative  moment $\mathtt{a}_i$ to each discrete ER and encode the contribution of the continuous family in a $2\times2$ PSD Hankel matrix $\hat{\mathcal{H}}(\mathtt{m})$, where the bi-variate moments are defined by $\mathtt{m}^{ab}=\int z_1^a z_2^b\dd\mu$.
The positivity cone is obtained from Eq~\eqref{eq:summomF}, which is now
\begin{align}
 &F_{M2}=-\tfrac{16}{3}\mathtt{a}_5+4\mathtt{m}^{11},~~~~~~F_{M3}=-\tfrac{32}{3}(\mathtt{a}_5+\mathtt{a}_6),\nonumber\\
 &F_{S0}=2(\mathtt{a}_1+\mathtt{a}_7+\mathtt{a}_8),~~~~~F_{S1}=-2\mathtt{a}_4-4\mathtt{a}_8+2\mathtt{m}^{02},\nonumber\\
 &F_{S2}=-2\mathtt{a}_1+2\mathtt{a}_4+2\mathtt{a}_8,~~F_{T8}=-2\mathtt{a}_2-\mathtt{a}_3+2\mathtt{m}^{20},\nonumber\\
 &F_{T9}=4(\mathtt{a}_2+\mathtt{a}_3),
 \label{eq:BHmap}
\end{align}
together with $\mathtt{a}_i\geq0$ and $\hat{\mathcal{H}}(\mathtt{m})\succeq0$. Since there are more moment variables (11) than Wilson coefficients (7), the final positivity bounds can be obtained by standard quantifier elimination:
\begin{align}
&F_{S0}\geq0,~~F_{T9}\geq0,~~F_{M3}\leq0,~~\tilde{F}_1\geq0,~~\tilde{F}_2\geq0,\nonumber\\
&F_{S0}+F_{S2}\geq0,~~F_{M3}/2-\hat{F}\leq F_{M2}\leq \hat{F} , \label{eq:BHexact}
\end{align}
where we have defined
\begin{equation}
\tilde{F}_1=F_{S0}+F_{S1}+F_{S2},~\tilde{F}_2=2F_{T8}+F_{T9},~\hat{F}=(2\tilde{F}_1\tilde{F}_2)^{\frac12} .\nonumber
\end{equation}
The nonlinear bound above correlates the mixed $BH$ coefficients with those in the pure $BB$ and $HH$ sectors. In particular, for $F_{M3}=0$, it implies $F_{M2}^2\leq2\tilde F_1\tilde F_2$, so a nonzero $F_{M2}$ requires nonzero contributions from both pure sectors.

\Section{The inverse problem}
\label{sec:inverse}
The EFT inverse problem aims to reconstruct possible UV interactions from a given set of low-energy coefficients \cite{Dawson:2020oco, Cepedello:2022pyx, Calisto:2026pvv}. Since such a reconstruction is generally not unique, further restrictions are usually needed to narrow down the possible UV realizations. However, when the EFT coefficients lie on or near discrete ERs of the amplitude cone, positivity bounds can sharply constrain, and in suitable cases uniquely determine, the UV particle types and their interactions~\cite{Zhang:2020jyn,Zhang:2021eeo}. In this section, we extend this by applying our moment framework to UV reconstruction for continuous families of ERs.

Specifically, when the EFT coefficients lie on a continuous ER family, the Hankel matrix $\hat{\mathcal H}$ becomes singular, as the positivity constraints arise from the PSD of $\hat{\mathcal H}$.
A singular $\hat{\mathcal H}$, denoted as $\hat{\mathcal{H}}_{\rm s}$, has a null vector $q_{\rm s}$, which means $\hat{\mathcal{H}}_{\rm s} \mathbf{q}_{\rm s} =0$. Then, the integral representation of $\hat{\mathcal H}_{\rm s}$ implies, when a non-trivial UV completion exists,
\begin{align}
0&=\mathbf{q}^T_{\rm s}\hat{\mathcal H}_{\rm s}\mathbf{q}_{\rm s}
=\int(\mathbf{q}_{\rm s}^T\mathbf{z}_{\rm s})^2\,\dd\mu(z) ~~\Rightarrow~~ \mathbf{q}_{\rm s}^T \mathbf{z}_{\rm s}=0 .
\label{eq:dualkernel}
\end{align}
Remember that the parameters $\mathbf z_{\rm s}=(z_{\rm s,1}, z_{\rm s,2},...)$ characterize how strongly different two-particle channels $(ij)$ couple to a given UV state $X$ relative to one another. They arise because several independent channels can couple to the same UV state, and their relative coupling strengths are not fixed by symmetry alone. The boundary null condition $\mathbf{q}_{\rm s}^T \bm z_{\rm s}=0$ further constrains these relative strengths, and, as we will see, provides the key guidance for constructing (partial) UV realizations of the EFT.

\Section{UV reconstruction: $B$-$H$} Now, let us illustrate how the UV reconstruction works for the continuous ER family from $B$-$H$ scattering.
To isolate the continuous ER family, we first impose the saturation of the linear bounds from discrete ERs: $F_{S0}=F_{S0}+F_{S2}=F_{T9}=F_{M3}=0$, which forces $\mathtt{a}_i$ in \eref{eq:BHmap} to vanish. For simplicity, we first consider the $B$-$H$ sector at $F_{M3}=0$, whose positivity bounds are described by a curved cone in the 3D space of $(F_{S1},F_{T8},F_{M2})$,
\begin{align}
\!\hat{\mathcal H}^{S}
=
\begin{pmatrix}
F_{T8}/2 & F_{M2}/4\\
F_{M2}/4 &  F_{S1}/2
\end{pmatrix}
\succeq0
& ~\Rightarrow~
\begin{matrix*}
 4F_{S1} F_{T8}\geq F_{M2}^2 \\
 F_{S1}, F_{T8}\geq0
\end{matrix*}.
\end{align}
We are interested in saturating the quadratic bound above to obtain the boundary null condition \eqref{eq:dualkernel}. On this curved boundary, $\hat{\mathcal H}^{S}$ has 
a null vector $\mathbf{q}_{\rm s}^{S}=(-2 F_{S1},F_{M2})^T$, and the boundary null condition becomes
\begin{equation}
\frac{z_1}{z_2} = \frac{F_{M2}}{2 F_{S1}}.
\end{equation}
Recall that $z_1$ and $z_2$ are possible couplings of two-particle states $BB$ and $HH$ to the same UV state, so this equation means that the ratio of these two UV couplings can be probed by measuring the EFT coefficients $ F_{S1}$ and $F_{M2}$.

The parametrization of this continuous ER family also guides us to construct the relevant UV model for an EFT lying on this curved boundary. For a tree-level UV realization, the interacting part of the UV Lagrangian for this case is given by
\begin{equation}
\Lag_{\rm int}^{\Phi}
=
\frac{\Phi}{f}\left[
g_H(D_\mu H)^\dagger D^\mu H +
g_B B_{\mu\nu}B^{\mu\nu}
\right],
\label{eq:scalarUV0}
\end{equation}
where $f$ denotes the suppression scale of the dimension-5 interactions, $\Phi$ is a heavy scalar singlet with mass $M_\Phi$, $B_{\mu\nu}$ is the hypercharge field-strength tensor, and $D_\mu$ is the covariant derivative. The two operators in \eref{eq:scalarUV0} are dimension-5, so it is a partial UV completion and will be fully completed at a higher energy. In fact, as shown in Appendix~\ref{app:treeRe}, this is generic for tree-level matching involving $M$- and $T$-type aQGC operators, which can not be completed by weakly-coupled renormalizable theories.
Integrating out $\Phi$ at tree level gives
\begin{equation}
F_{S1}=\frac{g_H^2}{2f^2M_\Phi^2},
~~
F_{T8}=\frac{8g_B^2}{f^2M_\Phi^2},
~~
F_{M2}=-\frac{4g_Hg_B}{f^2M_\Phi^2},
\end{equation}
with all other dimension-8 aQGC coefficients vanishing at this order. As expected, this saturates the nonlinear positivity bound $F_{M2}^2=4F_{S1} F_{T8}$.
% The amplitudes $BB\to\Phi$ and $HH^\dagger\to\Phi$ are $(-4g_B,g_H)M_{\Phi}^2$,
After factoring out the kinematic and tensor structures, the amplitudes for $BB\to\Phi$ and $HH^\dagger\to\Phi$ yield the coefficient pair $(-4g_BM_\Phi^2,g_HM_\Phi^2)$,
which is proportional to both $(F_{M2},2 F_{S1})$ and the coupling vector $(z_1,z_2)$. This confirms that the Wilson coefficient vector from the UV model \eqref{eq:scalarUV0} lies on the continuous ER family specified by $\mathbf{e}_{\rm BH}(-2g_B/(fM_\Phi),g_H/(2fM_\Phi))$.

The same reasoning applies to the quadratic positivity bound involving only $M$-type aQGC coefficients in scattering involving $W$,$B$ and $H$. As detailed in Appendix~\ref{app:aQGCbound}, in this case, Hankel positivity gives
\begin{align}
\hat{\mathcal H}^{V}
\!=\!
\begin{pmatrix}
\tilde F_{13} & -F_{M5}\\
-F_{M5} & -4F_{M3}
\end{pmatrix}
\succeq0
\Rightarrow\!\!
\begin{matrix*}
-4F_{M3}\tilde F_{13}\geq F_{M5}^2 \\
-F_{M3}\geq0,\tilde F_{13}\geq0  
\end{matrix*}
\label{eq:Mbound}
\end{align}
where $\tilde F_{13}\equiv F_{M7}-8F_{M1}$.
The linear bounds derived from discrete ERs now do not require certain coefficients to vanish.
On the curved boundary above, the null vector is $\mathbf{q}_{\rm s}^{V}=(F_{M5},\tilde F_{13})^T$, which links the following coupling ratio of the $WH$ and $B$-$H$ channels to the IR coefficients
\begin{equation}
\frac{z_1}{z_2} = -\frac{\tilde F_{13}}{F_{M5}}.
\end{equation}
To realize saturation of the quadratic bound in Eq.~\eqref{eq:Mbound}, we introduce a massive complex vector $V_\mu$ transforming as an $SU(2)_L$ doublet with hypercharge $Y=1/2$.
The interaction Lagrangian takes the form
\begin{equation}
\!\!\Lag_{\rm int}^{V}
=\frac{V_\mu^\dagger}{f}
\left[ \tilde{g}_1\frac{\tau^I}{2} W^{I,\mu\nu}D_\nu H + \tilde{g}_2 B^{\mu\nu}D_\nu H
\right] +\hc,
\label{eq:dim5WBHL}
\end{equation}
where $\tilde{g}_1$ and $\tilde{g}_2$ are real couplings, $\tau^I$ are the Pauli matrices, and $W^{I,\mu\nu}$ is the $SU(2)_L$ field-strength tensor.
Integrating out $V_\mu$ at tree level and matching onto the dimension-8 aQGC basis yields
\begin{align}
F_{M1}=\!-\!\lambda \tilde{g}_1^2,
F_{M3}=\!-4\lambda \tilde{g}_2^2,F_{M5}=8\lambda \tilde{g}_1\tilde{g}_2,F_{M7}=\!-4\lambda \tilde{g}_1^2\nonumber,
\end{align}
where $\lambda=1/(f^2M_V^2)$. Indeed, this saturates the nonlinear bound $F_{M5}^2+4F_{M3}\tilde F_{13}=0$ and reproduces the continuous ER family via $\mathbf{e}_{18}(0,-2\tilde{g}_2,0,\tilde{g}_1)$.

In the $W$-$B$-$H$ case, we see that vector-doublet exchange can generate a negative contribution to $F_{M3}$. Therefore, to extend the $B$-$H$ analysis to $F_{M3}<0$, we must saturate the two Hankel matrices simultaneously:
\begin{align}
\hat{\mathcal H}_{\rm s}^{S}
&=
\begin{pmatrix}
\tilde F_2/4 & (F_{M2}+\tilde{\mathtt a}_5)/4\\
(F_{M2}+\tilde{\mathtt a}_5)/4 & \tilde F_1/2
\end{pmatrix},
\nonumber\\
\hat{\mathcal H}_{\rm s}^{V}
&=
\begin{pmatrix}
\tilde{\mathtt a}_5 & 0\\
0 & -F_{M3}/2-\tilde{\mathtt a}_5
\end{pmatrix},
\end{align}
where $\tilde{\mathtt a}_5=16\mathtt a_5/3$, with $\mathtt a_5$ associated with the last discrete ER of \eref{eq:BHERs}. Saturation of $\hat{\mathcal H}_{\rm s}^{V}$ implies either $\tilde{\mathtt a}_5=0$ or $\tilde{\mathtt a}_5=-F_{M3}/2$, which upon substituting into $\hat{\mathcal H}_{\rm s}^{S}$ gives the last bound in \eref{eq:BHexact}. Also, extracting the two null vectors of $\hat{\mathcal H}_{\rm s}^{S}$, we find that the two branches of the relative coupling ratio are linked to the EFT coefficients as follows
\begin{equation}
\frac{z_1}{z_2}
=
\pm\sqrt{\frac{\tilde F_2}{2\tilde F_1}}.
\end{equation}
On the other hand, the null vectors of $\hat{\mathcal H}_{\rm s}^{V}$ are $(1,0)^T$ and $(0,1)^T$, each corresponding to switching on one UV-vector exchange channel.

\begin{figure}
    \centering
    \includegraphics[width=0.8\linewidth]{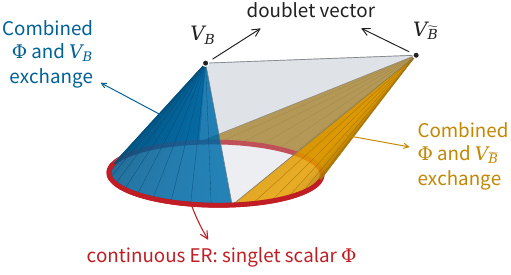}
    \caption{Section of the $BH$ positivity cone in $(\tilde F_1,\tilde F_2,F_{M2})$ coordinates, with  $F_{S0}=F_{S2}=F_{T9}=0$ and $\tilde F_1+\tilde F_2/2-F_{M3}/4=1$. The blue and orange surfaces saturate quadratic positivity bounds.
    The red curve represents the continuous ER family realized by a UV scalar singlet $\Phi$, while points $V_B$ and $V_{\tilde{B}}$ denote the discrete ERs realized by a UV vector doublet.}
    \label{fig:bhboundary}
\end{figure}

This motivates combining the scalar singlet with a vector doublet to realize the two saturation branches of the $BH$ bound at $F_{M3}<0$.
The combined interaction Lagrangian takes the form
\begin{align}
\Lag_{\rm int}
&=
\frac{\Phi}{f}\left[
g_H(D_\mu H)^\dagger D^\mu H
+
g_B B_{\mu\nu}B^{\mu\nu}
\right]
\nonumber\\
&\quad+
\left[
\frac{V_\mu^\dagger}{f}
\left(
\tilde{g}_3\tilde B^{\mu\nu}D_\nu H
+
\tilde{g}_4 B^{\mu\nu}D_\nu H
\right)
+\hc
\right]\nonumber,
\end{align}
where $\tilde B^{\mu\nu}=\epsilon^{\mu\nu\rho\sigma}B_{\rho\sigma}/2$, and $\tilde{g}_3,\tilde{g}_4$ are real couplings.
Guided by the null condition of the Hankel matrix $\hat{\mathcal{H}}^V_{\rm s}$, the tree-level EFT matching with $g_Hg_B<0$ in the first branch and $g_Hg_B>0$ in the second yields 
\begin{equation}
\begin{matrix*}
    F_{S1}=\frac{g_H^2}{2f^2M_\Phi^2}\\
    F_{T8}=\frac{8g_B^2}{f^2M_\Phi^2}
\end{matrix*}
~+~
\begin{cases}
    F_{M2}=-\frac{4g_Hg_B}{f^2M_\Phi^2},F_{M3}=-\frac{4\tilde{g}_4^2}{f^2M_V^2}, \\
    F_{M2}=-\frac{4g_Hg_B}{f^2M_\Phi^2}-\frac{2\tilde{g}_3^2}{f^2M_V^2},F_{M3}\!=\!-\frac{4\tilde{g}_3^2}{f^2M_V^2}
\end{cases}\nonumber
\end{equation}
where the two branches after ``$\{$'' correspond to turning on one vector coupling at a time. They correspond, respectively, to the following two combinations of a continuous ER family and a discrete ER:
\begin{align}
&\!\!\mathbf{e}_{\rm BH}\Big(\!-\!\frac{2g_B}{fM_\Phi},\frac{g_H}{2fM_\Phi}\Big)\!+\frac{4\tilde{g}_4^2}{f^2\!M_V^2}(0,-1,0,0,0,0,0)
\\
&\!\!\mathbf{e}_{\rm BH}\Big(\!-\!\frac{2g_B}{fM_\Phi},\frac{g_H}{2fM_\Phi}\Big)\!+\frac{2\tilde{g}_3^2}{f^2\! M_V^2}(-1,-2,0,0,0,0,0).
\end{align}
These ERs and their UV implications are illustrated in Figure \ref{fig:bhboundary}.

These examples illustrate how the continuous ER families and the associated boundary null conditions can guide the construction of candidate UV realizations.
We present further SMEFT examples of low-energy boundary saturation and the associated high-energy UV reconstruction in Appendix \ref{app:inverse}, covering a linear bound realized by heavy vector-doublet exchange, quadratic bounds involving aQGC coefficients of different types, and a one-loop example. 
For EFT coefficients near such boundaries, the relevant UV realizations are still dominated by these extremal contributions, and in many cases one may still single out a particular UV model, as in the case of discrete ERs~\cite{Zhang:2020jyn}. This is particularly useful for identifying the underlying UV model from experimental data, where measurement uncertainties make exact saturation of positivity bounds unlikely.

\Section{Conclusion}
We have applied the moment-problem approach to derive analytical positivity bounds from ERs of the amplitude cone, including the aQGC coefficients in the SMEFT. The new method gives us an exact handle on the curved boundaries of the positivity cone. Beyond constraining the EFT parameter space, our framework establishes a connection between the geometry of positivity bounds and the structure of UV interactions. In particular, the saturation of nonlinear positivity bounds can constrain how UV states couple to different scattering channels, providing guidance for reverse-engineering possible UV realizations from the continuous families of ERs. Future work can extend this framework to other scattering sectors and incorporate the resulting bounds into experimental fits to constrain new-physics models phenomenologically.

\acknowledgments
\section{Acknowledgments}
We would like to thank Zhuo-Yan Chen for helpful discussions. SYZ acknowledges
support from the National Natural Science Foundation of China under grant No. 12475074 and No. 12247103. This research is also supported by the advanced computing resources provided by the Supercomputing Center of the
USTC.

\bibliography{ref}

%\clearpage
\appendix
\setcounter{secnumdepth}{1}
\section{Analytic bounds for aQGCs in SMEFT}\label{app:aQGCbound}
This appendix presents the moment formulation of the full $W,B,H$ system in electroweak boson scattering, with the relevant interactions parameterized by $22$ aQGC operators.
These $22$ real coefficients comprise $3$ pure-Higgs ($F_{S\mathcal{I}_1}$), $9$ mixed gauge-boson--Higgs ($F_{M\mathcal{I}_2}$), and $10$ pure gauge-boson ($F_{T\mathcal{I}_3}$) coefficients (see Appendix \ref{app:aQGCs}).
We first express these coefficients in terms of moment variables and then collect the analytic positivity bounds obtained from the PSD conditions on the moment matrices.

Following the structure of the ER families (see Appendix~\ref{app:PERaQGC}), let $\mathtt{a}_i$ ($i=1,\ldots,8$) denote the zero-variate moments associated with the isolated ERs.
For the two-, three-, and four-parameter ER families, we denote the moments by $\mathtt{b}_j^{mn}$ ($j=1,\ldots,5$), $\mathtt{c}_{1,2}^{mnl}$, and $\mathtt{d}^{mnlk}$, respectively.
Let us present an example to illustrate the moment formulation of the aQGC coefficients from the ERs. 
Within the positivity cone, the amplitude vector can be expressed as a non-negative linear combination of these ERs.
For example, the $F_{M0}$ component receives contributions from $\vec{e}_{13},\vec{e}_{17}$ and $\vec{e}_{18}$.
Using \eref{eq:summomF}, we obtain
\begin{align}
F_{M0}&=\!-\frac{8}{3}\!\int x^2\dd \mu_1(x)+2\!\int  yz \dd\mu_2(y,z) \!-\!\frac{8}{3}\!\int x^2 \dd\mu_3(x)\nonumber\\
&=-\frac{8}{3}\mathtt{b}_3^{20}+2\mathtt{c}_2^{011}-\frac{8}{3}\mathtt{d}^{2000}.
\end{align}
Once the coefficient--moment relations are established, we simplify the PSD constraints using chordal completion.
For example, consider a $4\times4$ moment block with an independently free off-diagonal entry $\tilde{\mathtt d}=\mathtt{d}^{0110}$.
Chordal PSD completion gives
\begin{align}
 \!\!\exists\,\tilde{\mathtt d}:\;
 \!\!\hat{\mathcal{H}}\equiv \begin{pmatrix}
 A&\bm u&\bm v\\
 \bm u^T&a&\tilde{\mathtt d}\\
 \bm v^T&\tilde{\mathtt d}&b
 \end{pmatrix}
 \succeq0
 \Longleftrightarrow
 \begin{cases}
 \begin{pmatrix}
 A&\bm u\\
 \bm u^T&a
 \end{pmatrix}\succeq0\\
 \begin{pmatrix}
 A&\bm v\\
 \bm v^T&b
 \end{pmatrix}\succeq0
 \end{cases} ,\nonumber
% \label{eq:chordal}
\end{align}
where $A$ is the common $2\times2$ principal block and $\bm u,\bm v$ are two-component vectors.
Since $\tilde{\mathtt d}$ enters neither the coefficient--moment relations nor any other constraints, this equivalence eliminates it by replacing the original PSD condition with two smaller PSD conditions.
The resulting aQGC system involves $43$ moment variables constrained by the PSD of the Hankel matrices, which are at most $3\times3$. 
Then we can derive the analytic bounds using the elimination procedure described in the main text.
For compactness, we define the following linear combinations:
\begin{align}
 &\tilde{F}_{1}=F_{S0}+F_{S1}+F_{S2},~~~~~~\tilde{F}_{2}=2F_{T8}+F_{T9},\nonumber\\
 &\tilde{F}_{3}=8F_{T0}+4F_{T1}+3F_{T2}+2F_{T3},\nonumber\\
 &\tilde{F}_{4}=4F_{T1}+F_{T2}+2F_{T3},~~~\tilde{F}_{5}=F_{T2}+2F_{T3},\nonumber\\
 &\tilde{F}_{6}=4F_{T0}+4F_{T1}+3F_{T2}+5F_{T3},\nonumber\\
 &\tilde{F}_{7}=12F_{T0}+4F_{T1}+5F_{T2}+3F_{T3},\nonumber\\
 &\tilde{F}_{8}=12F_{T0}+4F_{T1}+6F_{T2}+5F_{T3},\nonumber\\
&\tilde{F}_{9}\!=\!2F_{T4}+4F_{T6}+F_{T7},~\tilde{F}_{10}\!=\!2F_{T0}+2F_{T1}+F_{T2}+F_{T3},\nonumber\\
 &\tilde{F}_{11}=6F_{T0}+2F_{T1}+4F_{T2}+3F_{T3},\nonumber\\
 &\tilde{F}_{12}=F_{S0}+F_{S2},~~~~~~~~\tilde{F}_{13}=F_{M7}-8F_{M1},\nonumber\\
 &\bar{F}_1=\max\{12F_{T5}+F_{T7}+F_{T4},12F_{T5}+F_{T7}-F_{T4}, \nonumber\\
 &\!-\!12F_{T5}\!-\!4F_{T6}\!-\!5F_{T7}\!-\!3F_{T4},\!12F_{T5}\!-\!4F_{T6}\!+\!F_{T7}\!-\!F_{T4}\}\nonumber\\
 &\bar{F}_2=\max(F_{T4}+4F_{T5},-F_{T4}-4F_{T5}-2F_{T7}) \nonumber\\
 &\bar{F}_3=\max\{8F_{T5}+F_{T7},-8F_{T5}-3F_{T7}-2F_{T4}-4F_{T6}\} \nonumber\\
 &\bar{F}_4=\max(-F_{T7}-4F_{T6},-F_{T7}-2F_{T4}).
 \label{eq:shortfull}
\end{align}
The following linear and quadratic bounds are derived from the moment construction.
The linear bounds are
\begin{align}
&F_{T9}\geq0~\tilde{F}_{5}\geq0~F_{T7}\geq0~\tilde{F}_{6}\geq0~\tilde{F}_{3}\geq0~\tilde{F}_{7}\geq0\nonumber\\
&\tilde{F}_{4}\geq0~F_{T2}\geq0~\tilde{F}_{2}\geq0~\tilde{F}_{9}\geq0~\tilde{F}_{1}\geq0~\tilde{F}_{12}\geq0\label{eq:linbound}\\
&F_{S0}\geq0~F_{M3}\leq0~\tilde{F}_{13}-2F_{M9}\geq0~\tilde{F}_{13}+2F_{M9}\geq0.\nonumber
\end{align}
The nine quadratic bounds involving only T-type Wilson coefficients are
\begin{align}
&16\tilde{F}_{10}\tilde{F}_{2}\geq(F_{T7}+4F_{T5})^2,~~16\tilde{F}_{11}\tilde{F}_{2}\geq3(4F_{T5}+F_{T7})^2\nonumber\\
&4F_{T9}(F_{T2}+F_{T3})\geq F_{T4}^2,~~~2\sqrt{(12F_{T8}+7F_{T9})\tilde{F}_{7}}\geq \bar{F}_1\nonumber\\
&2\sqrt{(4F_{T8}+3F_{T9})\tilde{F}_{8}}\geq\sqrt3\,\bar{F}_2,~~4\sqrt{\tilde{F}_{3}\tilde{F}_{2}}\geq \bar{F}_3 \label{eq:Tfull} \\
&4\sqrt{6\tilde{F}_{7}\tilde{F}_{2}}\geq3\bar{F}_3,~~2\sqrt{2\tilde{F}_{5}F_{T9}}\geq \bar{F}_4,~~4\sqrt{\tilde{F}_{7}F_{T9}}\geq \bar{F}_4 \nonumber
\end{align}
The linear conditions above ensure that all quantities under the square roots in ~\eref{eq:Tfull} are non-negative.
The quadratic bound involving only M-type coefficients is
\begin{equation}
  -4F_{M3}\tilde{F}_{13}\geq F_{M5}^2.
\end{equation}
Defining $\Delta_M\!\!=\!\!(-F_{M3}\tilde{F}_{13})^{1/2}$, the eleven quadratic bounds involving Wilson coefficients of different types are
\begin{align}
&2\sqrt{2\tilde{F}_{1}\tilde{F}_{2}}\geq\max(F_{M3}-2F_{M2},2F_{M2}),2\sqrt{2F_{S0}\tilde{F}_{4}}\geq F_{M1}\nonumber\\
&8\sqrt{2\tilde{F}_{1}\tilde{F}_{3}}\geq\max(\!-\!16F_{M0}\!-\!\tilde{F}_{13},16F_{M0}),\sqrt{2\tilde{F}_{12}\tilde{F}_{6}}\geq F_{M1} \nonumber\\
&2\Delta_M+8\sqrt{2F_{S0}F_{T7}}\geq|F_{M8}|,~2\sqrt{\tilde{F}_{12}\tilde{F}_{4}}\geq F_{M1} \label{eq:mixfull}\\
&2\sqrt{\tilde{F}_{12}\tilde{F}_{5}}\geq F_{M1},~2\sqrt{2F_{S0}\tilde{F}_{5}}\geq F_{M1},~2\sqrt{F_{S0}\tilde{F}_{6}}\geq F_{M1}\nonumber\\
&2\sqrt{\tilde{F}_{12}\tilde{F}_{9}}+\Delta_M\geq|F_{M4}|,~2\sqrt{2\tilde{F}_{1} \tilde{F}_{9}}+\Delta_M\geq|F_{M4}|\nonumber
\end{align}

\section{Additional examples of UV reconstruction}
\label{app:inverse}
In this appendix, we provide additional aQGC examples in the SMEFT to illustrate how to reconstruct UV models that saturate positivity bounds with curved boundaries.

\Section{Tree-level UV for $W$-$H$}
Consider the quadratic bounds in the second line of \eref{eq:mixfull}: $8(2\tilde{F}_{1}\tilde{F}_{3})^{1/2}\geq\max(\!-\!16F_{M0}\!-\!\tilde{F}_{13},16F_{M0})$, which involve $W$ and Higgs bosons.
When $\tilde{F}_{13}=0$, the saturation of these bounds gives $2F_{M0}^2=\tilde{F}_1\tilde{F}_3$, making the corresponding Hankel block singular: 
\begin{equation}
  \hat{\mathcal{H}}_{\rm s}=
  \begin{pmatrix}
  2\tilde{F}_1 & F_{M0}/2\\
  F_{M0}/2 &\tilde{F}_3/16
  \end{pmatrix}
\end{equation}
The boundary null condition then fixes the UV coupling ratio to be $(z_1,z_2)\propto(4\tilde{F}_1,F_{M0})$, where $z_1$ and $z_2$  are possible couplings of two-particle states $HH$ and $WW$ to the same UV state.

When $\tilde{F}_{13}>0$, the quadratic bounds can be rewritten as $\!-\!\tilde{F}_{13}/8\!-\!(2\tilde{F}_{1}\tilde{F}_{3})^{1/2}\leq 2F_{M0}\leq (2\tilde{F}_{1}\tilde{F}_{3})^{1/2}$. The singular Hankel matrices on the boundary are
\begin{align}
\hat{\mathcal H}^{S}_{\rm s}(\tilde{\mathtt d})
&=
\begin{pmatrix}
2\tilde F_1&(F_{M0}+\tilde{\mathtt d})/2\\
(F_{M0}+\tilde{\mathtt d})/2&\tilde F_3/16
\end{pmatrix}\succeq0,
\\
\hat{\mathcal H}^V_{\rm s}(\tilde{\mathtt d})
&=
\begin{pmatrix}
\tilde F_{13}/16-\tilde{\mathtt d}&F_{M9}/16\\
F_{M9}/16&\tilde{\mathtt d}
\end{pmatrix}\succeq0,
\end{align}
where $\!\tilde{\mathtt d}\!=\!8\mathtt d^{200}/3$.
The positive diagonal entries of $\hat{\mathcal{H}}^V_{\rm s}$ enforce $0\leq \tilde{\mathtt d}\leq\tilde F_{13}/16$. The upper and lower saturation branches fix $\tilde{\mathtt d}=0$ and $\tilde{\mathtt d}=\tilde F_{13}/16$, respectively.
For $\tilde F_1,\tilde F_3>0$, saturation eliminates the auxiliary diagonal entries and makes the Hankel block $\hat{\mathcal{H}}^S_{\rm s}$ rank one, with the null vector ${\bf q}_{\rm s}=
(-(F_{M0}+\tilde{\mathtt d}), 4\tilde F_1)^T$.
Its boundary null condition then implies that the UV coupling ratio can be probed by the EFT coefficients as follows
\begin{equation}
\frac{z_2}{z_1}
=\frac{F_{M0}+\tilde{\mathtt d}}{4\tilde F_1}
=\pm\sqrt{\frac{\tilde F_3}{32\tilde F_1}},\nonumber
\end{equation}
with the positive and negative signs corresponding to the upper and lower branches of $\tilde{\mathtt d}$, respectively.
On these branches, the nonnegativity of the $2\times2$ principal minor of $\hat{\mathcal{H}}^V_{\rm s}$ also requires $F_{M9}=0$, and the corresponding  null vectors are $(0,1)$ and $(1,0)$.
These channel-coupling relations guide the UV reconstruction below.

For the corresponding UV model, when $\tilde{F}_{13}=0$, we are led to consider an $SU(2)_L$-singlet scalar $S$ of mass $M_S$ with the interaction
\begin{align}
 \Lag^S_{\rm int}=\frac{S}{f}\left[g_H(D_\mu H)^\dagger D^\mu H
 +g_W W^I_{\mu\nu}W^{I\mu\nu}\right].
\end{align}
To see this, integrating out $\!S\!$ at tree level yields 
\begin{equation}
F_{S1}=\frac{g_H^2}{2f^2M_S^2}, ~~F_{M0}=-\frac{2g_Hg_W}{f^2M_S^2},~~ F_{T0}=\frac{2g_W^2}{f^2M_S^2},
\end{equation}
which reproduce the quadratic bound $F_{M0}^2=4F_{S1}F_{T0}=\tilde{F}_{1}\tilde{F}_{3}/2$. For this scalar model, explicit matching identifies $(-g_W,g_H)\propto(F_{M0},4\tilde{F}_1)$. 

We extend this construction to $\tilde{F}_{13}>0$ by adding a vector doublet, which contributes positively to $\tilde{F}_{13}$.
In this case, we are led to consider the combined interaction,
\begin{align}
  \Lag^{SV}_{\rm int}&=\frac{S}{f}(g_H(D_\mu H)^\dagger D^\mu H
 +g_W W^I_{\mu\nu}W^{I\mu\nu})\nonumber\\
 &~~~~+(g_V V^\dagger_\mu(\hat{c}_1 J^\mu-i\hat{c}_2 \tilde{J}^\mu)+\hc)
\end{align}
where $J^\mu\equiv\tau^I/2\,W^{I,\mu\nu}D_\nu H, ~ \tilde J^\mu\equiv\tau^I/2\,\tilde W^{I,\mu\nu}D_\nu H$ and $\tilde W_{\mu\nu}\equiv\epsilon_{\mu\nu\rho\sigma}W^{\rho\sigma}/2$.
The boundary null conditions of $\hat{\mathcal{H}}^V_{\rm s}$ select $(\hat{c}_1,\hat{c}_2)=(1,0)$ on the upper branch and $(\hat{c}_1,\hat{c}_2)=(0,1)$ on the lower branch. Both branches retain the scalar-exchange coefficients $F_{S1}=g_H^2/(2f^2M_S^2)$ and $F_{T0}=2g_W^2/(f^2M_S^2)$. The M-type Wilson coefficients in the upper and lower branches are, respectively,
\begin{align}
\!\!&F_{M0}=-2\frac{g_Hg_W}{f^2M_S^2},~F_{M1}=-\frac{|g_V|^2}{f^2M_V^2},~F_{M7}=-4\frac{|g_V|^2}{f^2M_V^2} .\nonumber\\
&F_{M0}=-2\frac{g_Hg_W}{f^2M_S^2}-\frac{|g_V|^2}{4f^2M_V^2},~F_{M1}=0,~F_{M7}=4\frac{|g_V|^2}{f^2M_V^2}\nonumber .
\end{align}
Both saturate the corresponding branches of the positivity bound at tree level, with the relative couplings given by $g_W/g_H=\pm\tilde{F}_3^{1/2}/(32\tilde{F}_1)^{1/2}$.

\Section{Loop UV completion for M-operators}
As a complementary example, we consider a renormalizable model that generates the M-type aQGC coefficients at one loop. The model contains a vectorlike $SU(2)_L$-doublet fermion $\Psi$ and a singlet fermion $N$, with the Lagrangian
\begin{align}
    \Lag_{\rm f}\!=\!\bar\Psi(i\slashed{D}\!-\!M_\Psi)\Psi
+\bar N(i\slashed{D}\!-\!M_N)N\!-\!c_y(\bar\Psi HN\!+\!\bar NH^\dagger\Psi) \nonumber,
\end{align}
where $c_y$ is a real Yukawa coupling.
The Yukawa and gauge interactions in $\Lag_{\rm f}$ generate the transitions $WH\to\Psi\bar N$ and $BH\to\Psi\bar N$.
Matching the tree-level EFT amplitudes with the one-loop UV amplitudes, we find the following aQGC coefficients 
\begin{align}
  &F_{M0}=-\frac{g^2 K}{8}[f(r)-l(r)],~~F_{M1}=-\frac{g^2K}{2}[f(r)+l(r)],\nonumber\\
  &F_{M2}=-\frac{g'^2 K}{4}[f(r)-l(r)],~F_{M3}=-g'^2Kf(r),\nonumber\\
  &F_{M4}=-gg'K[f(r)-l(r)],~F_{M5}=4gg'Kf(r),\nonumber\\
  &F_{M7}=-4g^2Kl(r), \label{eq:loopmatching}
\end{align}
where $K\equiv c_y^2/(4\pi^2M_\Psi^4)$, $r\equiv M_N/M_\Psi$, and the loop functions satisfy $f(r)>|l(r)|$, whose expressions are cumbersome and not needed for the present discussion. The resulting coefficients also admit a representation as a non-negative combination of ERs: $
  \mathbf{F}=\mathbf{e}_{18}(x_+,y_+,y_-,x_-)+\mathbf{e}_{18}(-x_+,y_+,-y_-,x_-)$
where $4 x_\pm=g[K(f(r)\mp l(r))]^{1/2},4y_\pm=-g'[K(f(r)\pm l(r))]^{1/2}$.
Consequently, the quadratic positivity bound in \eref{eq:Mbound} $-4F_{M3}\tilde F_{13}\geq F_{M5}^2$ can also be saturated at one-loop order. 
In particular, the saturated coefficients give $(g',2g)\propto(F_{M5},2\tilde F_{13})$.
Unlike the single-particle exchanges considered there, the present one-loop contribution  additionally contains another ER.

\section{Tree-level selection rules for the M- and T-sectors}
\label{app:treeRe}

The partial UV completions constructed in this paper generate $M$- and $T$-type aQGC contributions through dimension-5 interactions or loops of heavy particles. One may wonder whether $M$- and $T$-type aQGCs can instead arise at tree level in a weakly coupled renormalizable UV theory. In this appendix, we show that this is not possible if we restrict heavy vector bosons to those arising from spontaneous gauge symmetry breaking, while allowing for generic scalars and fermions.

\begin{figure}[h]
\centering
\begin{minipage}[t]{0.48\linewidth}
  \centering
  \includegraphics[width=0.8\linewidth]{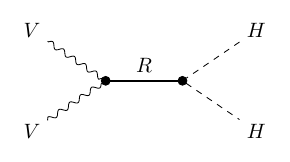}
  \par\small  $VVHH$: $RVV\times RHH$
\end{minipage}
\hfill
\begin{minipage}[t]{0.48\linewidth}
  \centering
  \includegraphics[width=0.8\linewidth]{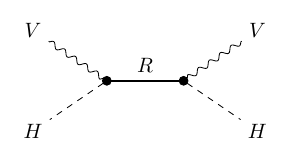}
  \par\small  $VVHH$: $RVH\times RVH$
\end{minipage}
\hfill
\begin{minipage}[t]{0.48\linewidth}
  \centering
  \includegraphics[width=0.8\linewidth]{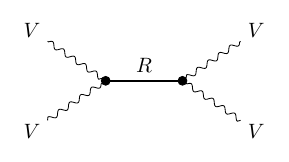}
  \par\small  $VVVV$: $RVV\times RVV$
\end{minipage}
\vspace{1em}
\caption{Tree-level four-point diagrams involving light gauge
bosons $V$, Higgs fields $H$, and heavy particles $R$.
Wavy, dashed, and thick solid lines denote $V$, $H$,
and $R$, respectively.
}
\label{fig:heavy_exchange}
\end{figure}

Note that the $M$- and $T$-type dimension-8 aQGC operators in the SMEFT have the schematic forms
\begin{equation}
    \mathcal O_M\sim F^2(DH)^\dagger DH,
    \quad
    \mathcal O_T\sim F^4,
\end{equation}
where $F$ denotes the gauge-boson field-strength tensor (Appendix \ref{app:aQGCs}). 
At tree level, the four-point EFT amplitudes from these operators are matched to four-point UV diagrams containing either one quartic vertex or two cubic vertices joined by a heavy propagator.
Since a renormalizable quartic bosonic vertex contains no derivatives, it cannot generate the fourth-order momentum dependence required for dimension-8 amplitudes. Therefore, at tree level, the $M$- and $T$-type aQGC coefficients must arise from heavy-particle exchange.
Denoting a light gauge boson by $V$ and a heavy particle by $R$, the $VVVV$ amplitude requires $RVV$ vertices, while the $VVHH$ amplitude requires either $RVV$ and $RHH$ vertices or two $RVH$ vertices, see Figure~\ref{fig:heavy_exchange}.
Since a fermion cannot couple to two bosons through a Lorentz-invariant cubic interaction, only heavy scalars and vectors need to be considered.

For the $RVV$ vertex, when $R$ is a heavy scalar, the gauge-invariant interactions first arise at dimension-5,
\begin{equation}
    RF_{\mu\nu}F^{\mu\nu},
    \qquad
    RF_{\mu\nu}\tilde F^{\mu\nu},
\end{equation}
and are therefore absent from the renormalizable theory.
Another possible source of the $RVV$ vertex is the renormalizable scalar kinetic term, expanded around a possible UV symmetry-breaking vacuum $v$. However, the resulting coupling is proportional to $t^av$, where $t^a$ is an unbroken electroweak gauge generator. Since the unbroken generators leave the vacuum invariant, $t^av=0$, and the vertex vanishes.
For a heavy gauge vector $R$, which is assumed to obtain its mass via UV spontaneous symmetry breaking, the $RVV$ vertex is proportional to the structure constant $f^{abA}$, where $a,b$ label light gauge bosons and $A$ labels $R$. Since the unbroken gauge generators form a closed algebra, $f^{abA}=0$, and the $RVV$ vertex vanishes.
Thus, the $RVV$ vertex is absent for both heavy scalars and gauge vectors.

For a heavy scalar, the candidate $RVH$ vertex arises from the scalar kinetic term and is proportional to $(t^a)_{RH}$, where $(t^a)_{RH}$ denotes the matrix element of $t^a$ between $R$ and $H$. Gauge invariance requires the scalar mass matrix to commute with $t^a$, which implies
\begin{equation}
    [\mathcal M_\Phi^2,t^a]=0
    ~~\Longrightarrow~~
    (M_R^2-m_H^2)(t^a)_{RH}=0.
\end{equation}
Since the heavy and light scalars have different masses, $(t^a)_{RH}=0$, and hence the $RVH$ vertex vanishes.
For a heavy gauge vector $R_\mu^A$, let $T^A$ denote its broken gauge generator and $t^a$ the unbroken generator associated with the light gauge boson $V_\mu^a$, which are antisymmetric. The vacuum $v$ satisfies $t^av=0$, while the physical Higgs field $h$ is orthogonal to the vectors $T^Av$. Expanding the scalar kinetic term, the $R_\mu^A V^{a\mu}h$ coupling is proportional to
\begin{align}
    \langle T^Av,t^ah\rangle
    &=-\langle t^aT^Av,h\rangle
    =-\langle[t^a,T^A]v,h\rangle=0
\end{align}
where $\langle,\rangle$ is the inner product.
The last equality follows because $[t^a,T^A]v$ is a linear combination of vectors $T^A v$, which are orthogonal to the physical Higgs field. Thus, the $RVH$ vertex vanishes.

Since neither the $RVV$ nor the $RVH$ vertex is allowed, the independent $M$- and $T$-type aQGC amplitudes receive no tree-level contributions,
\begin{equation}
F_{Mi}^{\rm tree}=F_{Ti}^{\rm tree}=0,
\end{equation}
in weakly coupled renormalizable UV theories whose heavy vector fields obtain their masses from spontaneous symmetry breaking. Nonzero contributions may instead arise from loop effects or dimension-5 interactions that we construct in this paper.

\newpage
\onecolumngrid
\section{aQGC operators in SMEFT}\label{app:aQGCs}
In this appendix, we list all dimension-8 aQGC operators for ease of reference:
\begin{align}
  O_{S,0} & =\left[  \left(  D_{\mu}\Phi \right)  ^{\dagger}D_{\nu}\Phi \right] \times \left[  \left(  D^{\mu}\Phi \right)  ^{\dagger}D^{\nu}\Phi \right], &\quad O_{M,9} & =-\frac{i}{2} D^{\mu}\Phi^{\dagger}\hat{W}_{\mu \rho}\tilde{W}^{\nu \rho}D_{\nu}\Phi+\mathrm{h.c.},\nn
  O_{S,1} & =\left[  \left(  D_{\mu}\Phi \right)  ^{\dagger}D^{\mu}\Phi \right] \times \left[  \left(  D_{\nu}\Phi \right)  ^{\dagger}D^{\nu}\Phi \right], &\quad  O_{T,0} & =\mathrm{Tr}\left[  \hat{W}_{\mu \nu}\hat{W}^{\mu \nu}\right]\mathrm{Tr}\left[  \hat{W}_{\alpha \beta}\hat{W}^{\alpha \beta}\right],\nn
  O_{S,2} & =\left[  \left(  D_{\mu}\Phi \right)  ^{\dagger}D_{\nu}\Phi \right] \times \left[  \left(  D^{\nu}\Phi \right)  ^{\dagger}D^{\mu}\Phi \right],&\quad O_{T,1} & =\mathrm{Tr}\left[  \hat{W}_{\alpha \nu}\hat{W}^{\mu \beta}\right] \mathrm{Tr}\left[  \hat{W}_{\mu \beta}\hat{W}^{\alpha \nu}\right],\nn
  O_{M,0} & =\mathrm{Tr}\left[  \hat{W}_{\mu \nu}\hat{W}^{\mu \nu}\right] \times \left[  \left(  D_{\beta}\Phi \right)  ^{\dagger}\left(  D^{\beta}
  \Phi \right)  \right]  ,&\quad O_{T,2} & =\mathrm{Tr}\left[  \hat{W}_{\alpha \mu}\hat{W}^{\mu \beta}\right]  \mathrm{Tr}\left[  \hat{W}_{\beta \nu}\hat{W}^{\nu \alpha}\right],\nn
  O_{M,1} & =\mathrm{Tr}\left[  \hat{W}_{\mu \nu}\hat{W}^{\nu \beta}\right]  \times \left[  \left(  D_{\beta}\Phi \right)  ^{\dagger}D^{\mu}\Phi \right] &\quad    O_{T,3} & =\mathrm{Tr}\left[ \hat{W}_{\mu\nu}\hat{W}_{\alpha\beta} \right]\mathrm{Tr}\left[ \hat{W}^{\alpha\nu}\hat{W}^{\mu\beta} \right],\nn
  O_{M,2} & =\hat{B}_{\mu \nu}\hat{B}^{\mu \nu}\times \left[  \left(
  D_{\beta}\Phi \right)  ^{\dagger}D^{\beta}\Phi \right]  ,&\quad  O_{T,4} & =\mathrm{Tr}\left[ \hat{W}_{\mu\nu}\hat{W}_{\alpha\beta} \right] \hat{B}^{\alpha \nu}\hat{B}^{\mu \beta},\nn
  O_{M,3} & =\hat{B}_{\mu \nu}\hat{B}^{\nu \beta}\times \left[  \left(  D_{\beta}\Phi \right)  ^{\dagger}D^{\mu}\Phi \right]   &\quad  O_{T,5} & =\mathrm{Tr}\left[  \hat{W}_{\mu \nu}\hat{W}^{\mu \nu}\right]\hat{B}_{\alpha \beta}\hat{B}^{\alpha \beta},\nn
  O_{M,4} & =\frac{1}{2}\left[  \left(  D_{\mu}\Phi \right)  ^{\dagger}\hat{W}_{\beta \nu}\left(  D^{\mu}\Phi \right)  \right]  \times \hat{B}^{\beta \nu},&\quad  O_{T,6} & =\mathrm{Tr}\left[\hat{W}_{\alpha \nu}\hat{W}^{\mu \beta}\right]  \hat{B}_{\mu \beta}\hat{B}^{\alpha \nu}, \nn
  O_{M,5} & =\frac{1}{4}\left[  \left(  D_{\mu}\Phi \right)  ^{\dagger}\hat{W}_{\beta \nu
  }\left(  D^{\nu}\Phi \right)  \right]  \times \hat{B}^{\beta \mu}+\mathrm{h.c.}, &\quad  O_{T,7} & =\mathrm{Tr}\left[  \hat{W}_{\alpha \mu}\hat{W}^{\mu \beta}\right]  \hat{B}_{\beta \nu}\hat{B}^{\nu \alpha},\nn
  O_{M,7} & =\frac{1}{4}\left[  \left(  D_{\mu}\Phi \right)  ^{\dagger}\hat{W}_{\beta \nu
  }\hat{W}^{\beta \mu}\left(  D^{\nu}\Phi \right)  \right]  ,&\quad  O_{T,8} &  =\hat{B}_{\mu \nu}\hat{B}^{\mu \nu}\hat{B}_{\alpha \beta}\hat{B}^{\alpha \beta},\nn
  O_{M,8} & =\frac{i}{4} \left[ D_{\mu}\Phi^{\dagger}\tilde{W}_{\nu \rho}D^{\nu}\Phi \right]  \hat{B}^{\mu \rho}+\mathrm{h.c.}, &\quad   O_{T,9} & =\hat{B}_{\alpha \mu}\hat{B}^{\mu \beta}\hat{B}_{\beta \nu}\hat{B}^{\nu \alpha}, 
\end{align}
where $\epsilon_{\mu\nu\rho\sigma}$ is the Levi-Civita tensor with $\epsilon_{0123}=1$, $D_\mu=\partial_\mu-ig W^I_\mu\tau^I/2-\frac{i}{2}g'B_\mu$
\begin{align}
  \hat{W}^{\mu\nu}&\equiv ig\frac{\tau^I}{2}W^{I,\mu\nu}, \quad \hat{B}^{\mu\nu}\equiv ig'\frac{1}{2}B^{\mu\nu},\quad \tilde{W}^{\mu\nu}\equiv ig\frac{\tau^I}{2}\left(\frac{1}{2}\epsilon_{\mu\nu\rho\sigma}W^{I,\rho\sigma}\right),\nonumber\\
  W^{I}_{\mu\nu}&=\partial_\mu W^I_{\nu}-\partial_\nu W^I_{\mu}+g\epsilon_{IJK}W^J_{\mu} W^K_{\nu},\quad B_{\mu\nu}=\partial_\mu B_\nu-\partial_\nu B_\mu,\label{eq:WBfield}
\end{align}
and $g$ and $g'$ are $SU(2)_L$ and $U(1)_Y$ gauge couplings, respectively. The effective Lagrangian for the aQGC interactions is written as
\begin{equation}
  \mathcal{L}_{\rm aQGC}=\sum_{i} \frac{f_{S,i}}{\Lambda^4}O_{S,i}+\sum_{i} \frac{f_{M,i}}{\Lambda^4}O_{M,i}+\sum_{i} \frac{f_{T,i}}{\Lambda^4}O_{T,i}.
\end{equation}
We will find it convenient to absorb the electromagnetic coupling constant $e$, as well as $\cos\theta_w$ and $\sin\theta_w$ of the Weinberg angle $\theta_w$, into the Wilson coefficients as follows
\begin{equation}
  F_{S,\mathcal{I}_1}\equiv f_{S,\mathcal{I}_1}, \quad F_{M,\mathcal{I}_2}\equiv \frac{e^2}{s_w^nc_w^{2-n}} f_{M,\mathcal{I}_2}, \quad F_{T,\mathcal{I}_3}\equiv \frac{e^4}{s_w^n c_w^{4-n}} f_{T,\mathcal{I}_3}.
\end{equation}
Here, $n=0$ for $\mathcal{I}_2=\{2,3\}$ and $\mathcal{I}_3=\{8,9\}$; $n=1$ for $\mathcal{I}_2=\{4,5,8\}$; $n=2$ for $\mathcal{I}_2=\{0,1,7,9\}$ and $\mathcal{I}_3=\{4,5,6,7\}$; and $n=4$ for $\mathcal{I}_3=\{0,1,2,3\}$.

\section{Extremal rays for aQGCs in SMEFT}
\label{app:PERaQGC}
In this appendix, we collect the extremal rays of the aQGC positivity cone in the SMEFT obtained in Ref~\cite{Chang:2026ztn}. These extremal rays provide the starting point for the moment formulation developed in this work. We reproduce their explicit expressions and parameterizations here for quick reference ($\vec{e}_6$ and $\vec{e}_9$ are not listed below as they arise from the direct group construction but are not truly extremal upon the $su$ crossing projection).
{\footnotesize
\begin{align}
    \vec{e}_1= &\left(0, 0, 0, 0, 0, 0, 0, 0, 0, 2, 0, -2, 0, 0, 0, 0, 0, 0, 0, 0, 0, 0, 0, 0, 0\right),\nn
    \vec{e}_2= &\left(0, 0, 0, 0, 0, 0, 0, 0, 0, 0, 0, 0, 0, 0, 0, 0, 0, -2 , -2 , 8 , 0, 0, 0, 0, 0\right),\nn
    \vec{e}_3= &\left(0,0,0,0,0,0,0,0,0,0,0,0,0,0,0,0,0,0,-2,8,0,0,0,\frac{1}{2},0\right),\nn
    \vec{e}_4= &\left(0,0,0,0,0,0,0,0,0,0,0,0,-\frac{1}{4},-\frac{1}{12},1,-\frac{1}{3},0,0,0,0,0,0,0,0,0\right),\nn
    \vec{e}_5= &\left(0,0,0,0,0,0,0,0,0,0,0,0,-\frac{1}{12},\frac{1}{4},0,0,0,0,0,0,0,0,0,0,0\right),\nn
    \vec{e}_7= &\left(0,0,0,0,0,0,0,0,0,0,0,0,-1,1,4,-4,0,0,0,0,0,0,1,0,0\right),\nn
    \vec{e}_8= &\left(0,0,0,0,0,0,0,0,0,0,0,0,0,0,0,0,-16 ,-4 ,4 ,16 ,0,0,0,1,0\right),\nn
    \vec{e}_{10}= &\left(0,0,0,0,0,0,0,0,0,0,-2,2,0,0,0,0,0,0,0,0,0,0,0,0,0\right),\nn
    \vec{e}_{11}(x,y)= &\left(0,0,0,0,0,0,0,0,0,0,0,0,0,-\frac{y^2}{2},0,y^2,4 x y,0,-2 x y,0,-2 x^2,4 x^2,0,0,0\right),\nn
    \vec{e}_{12}(x,y)= &\left(0,0,0,0,0,0,0,0,0,0,0,0,-\frac{y^2}{4},0,0,y^2,4 xy,-xy,0,0,-x^2,4 x^2,-\frac{y^2}{2},-\frac{xy}{2},0\right),\nn
    \vec{e}_{13}(x,y)= &\left(-\frac{8x^2}{3},-\frac{8}{3}\left(3 x^2+y^2\right),0,0,0,0,-\frac{64}{3}\left(x^2-y^2\right),0,\frac{64x y}{3},0,0,0,0,0,0,0,0,0,0,0,0,0,0,0,0\right),\nn
    \vec{e}_{14}(x,y)= &\left(0,0,0,0,0,0,0,0,0,0,0,0,0,0,4y^2,-4y^2,0,0,0,0,0,0,y^2,x^2,-x y\right),\nn
    \vec{e}_{15}(x,y)= &\left(0,0,0,0,-16 xy,0,0,0,0,0,-2 x^2,4 x^2,0,0,0,0,0,0,4y^2,0,0,0,0,0,0\right),\nn
    \vec{e}_{16}(x,y,z)= &\left(0,-8 y z,0,0,0,0,-64 y z,-64 x y,0,2 y^2,-4 y^2,2 y^2,0,0,0,0,-16 x^2,0,0,16 x^2,0,0,z^2,x^2,xz\right),\nn
    \vec{e}_{17}(x,y,z)= &\left(2 y z,0,4 x y,0,0,0,0,0,0,0,2 y^2,0,\frac{z^2}{2},0,0,0,0,2 xz,0,0,2 x^2,0,0,0,0\right),\nn    
    \vec{e}_{18}(x,y,u,z)= &\left(-x^2,-4z^2,-2 u^2,-4 (y^2+u^2),8 x u,-16 (z y+x u),16 (-z^2+x^2),16 (-x y+z u),-16 x z,0,\dots,0\right).\label{eq:potential ERs}
\end{align}
}
\noindent

\end{document}